\documentclass[twocolumn,showpacs,preprintnumbers,amsmath,amssymb,showkeys]{revtex4}

\usepackage{epsfig}
\usepackage{amssymb}

\newcommand\eqn[1]{(\ref{#1})}      
\newcommand\Eqn[1]{Eq.~(\ref{#1})}  
\newcommand\Fig[1]{Fig.~\ref{#1}}  
\newcommand\Ref[1]{Ref.~\cite{#1}}  

\newcommand{\beq}{\begin{equation}}
\newcommand{\eeq}{\end{equation}}
\newcommand{\ba}{\begin{array}}
\newcommand{\bea}{\begin{eqnarray}}
\newcommand{\ea}{\end{array}}
\newcommand{\eea}{\end{eqnarray}}

\begin{document}


\title{Effective potential for quantum scalar fields in a de Sitter geometry}

\author{Julien Serreau}
\affiliation{Astro-Particule et Cosmologie, Universit\'e Paris 7 -- Denis Diderot,\\ 10 rue A.~Domont et L.~Duquet, 75205 Paris cedex 13, France\footnote{APC is unit\'e mixte de recherche UMR7164 (CNRS, Universit\'e Paris 7, CEA, Observatoire de Paris).}}

\date{\today}

\begin{abstract}
We study the quantum theory of an $O(N)$ scalar field on de Sitter geometry at leading order in a nonpertative $1/N$-expansion. This resums the infinite series of so-called superdaisy loop diagrams. We obtain the de Sitter symmetric solutions of the corresponding, properly renormalized, dynamical field equations and compute the complete effective potential. Because of its self-interactions, the field acquires a strictly positive square mass which screens potential infrared divergences. Moreover, strongly enhanced ultralong-wavelength fluctuations prevent the existence of a spontaneously broken symmetry state in any dimension.
\end{abstract}

\maketitle

The impressive observational success of inflationary cosmology \cite{Peiris:2003ff, Parentani:2004ta} and the quest for high precision measurements of the cosmic microwave background (CMB) properties, such as the power spectrum, or the non-Gaussian correlations of temperature anisotropies \cite{Bartolo:2004if}, call for establishing a coherent quantum (field) theory (QFT) of inflation. Indeed, in this context, the large scale primordial density fluctuations imprinted on the CMB are of purely quantum origin and, as such, are subject to the whole complexity of quantum field dynamics. 


The issue of computing QFT radiative (loop) corrections in curved geometries has a long history \cite{BirellDavies}, which has been revived in recent years in the context of inflationary cosmology \cite{Boyanovsky:2005px,Weinberg:2005vy, Sloth:2006az}. In particular, the case of de Sitter (dS) geometry has attracted a lot of interest \cite{Prokopec:2002jn, Tsamis:2005hd, vanderMeulen:2007ah}, first because of its direct relevance to inflationary physics \cite{Parentani:2004ta} and second, because its high degree of symmetry makes it somewhat simpler \cite{Bros:1994dn}. It appears that perturbative calculations of radiative corrections of quantum field dynamics in expanding space-times (ST) share many features with similar calculations in flat ST at finite temperature -- due to the presence of an event horizon -- and out of equilibrium -- due to the cosmic expansion.

For instance, loop diagrams in cosmological STs exhibit secular terms, which grow unbounded with the number of $e$-folds \cite{Weinberg:2005vy, Tsamis:2005hd}. Those are well-known generic features of perturbative calculations for nonequilibrium systems \cite{Berges:2004vw}. Another standard problem of perturbation theory is the occurrence of infrared (IR) divergences for massless fields minimally coupled to gravity, relevant for inflation, akin to what happens for light bosonic fields at high temperatures or near a second order phase transition. Such singularities signal deep deficiencies of the approximation scheme and need to be summed \cite{Burgess:2009bs}. 

Efficient resummation techniques have been developed over the years to deal with these issues in equilibrium and nonequilibrium QFT in flat ST. Prominent examples include the dynamical renormalization group (DRG) \cite{Boyanovsky:1998aa}, or two-particle-irreducible (2PI) functional techniques \cite{Blaizot:2001nr, Berges:2004vw}. First attempts to apply such methods to the cosmological case include the work of Riotto and Sloth \cite{Riotto:2008mv} using 2PI and large-$N$ techniques \cite{Ramsey:1997qc} but eventually relying on the effective stochastic approach of \Ref{Starobinsky:1994bd}, or the DRG analysis of \Ref{Burgess:2009bs}. One of the main results of these works is the fact that, as was discovered earlier by Starobinsky and Yokoyama in the stochastic approach \cite{Starobinsky:1994bd}, the self-interacting scalar field acquires a curvature-induced effective mass, which prevents possible IR divergences. This is analog to the temperature-induced mass in thermal QFT \cite{Blaizot:2001nr}.

Here, we study an $O(N)$ scalar field theory on dS in the large-$N$ limit from the basic principles of QFT. This resums the infinite series of so-called superdaisy (or cactus) loop diagrams and is known to resum high temperature IR divergences as well as to successfully describe the phase structure of $O(N)$ scalar theories in flat ST \cite{Coleman:1974jh}. We solve the theory in the dS symmetric state consistent with basic renormalization requirements, the so-called Bunch-Davies vacuum, study the phase structure of the theory and compute the effective potential in $3+1$ dimensions \cite{Anderson:1985hz}. A more extended treatment, including the case of a discrete $Z_2$ symmetry ($N=1$) will presented elsewhere \cite{PS}.


We consider a $O(N)$-symmetric scalar field theory with classical action (a sum over $a=1,\ldots,N$ is implied)
\beq
\label{eq:action}
 {\cal S}[\varphi]=-\int_x\left\{{1\over2}\varphi_a\left(\square+m^2+\xi{\cal R}\right)\varphi_a+\frac{\lambda}{4!N}\left(\varphi_a\varphi_a\right)^2\right\},
\eeq
with the invariant measure $\int_x\equiv\int d^{d+1}x\,\sqrt{-g}$, on a $d+1$-dimensional expanding dS background with flat spatial sections. In terms of comoving spatial coordinates $\vec x$ and conformal time $-\infty<\eta<0$, the line element reads
\beq
 ds^2=a^2(\eta)\left(d\eta^2-\vec{dx}\cdot\vec{dx}\right),
\eeq
where $a(\eta)=-1/H\eta$. Here, ${\cal R}=d(d+1)H^2$ and $\square= H^2\eta^2(\partial_\eta^2-(d-1){\eta^{-1}}\partial_\eta+\nabla_x^2)$.

In the large-$N$ limit, the field equation reads \cite{Boyanovsky:1997cr, Ramsey:1997qc}
\beq
\label{eq:field}
 \left[\square+M^2(x)\right]\varphi_a(x)=0
\eeq
where the self-consistent square mass is given by \footnote{Our definition of $M^2$ includes the coupling to curvature.}
\beq
 \label{eq:mass}
 M^2(x)=m^2+d(d+1)\xi H^2+\frac{\lambda}{6}\phi^2(x)+\frac{\lambda}{6}F(x,x),
\eeq
in terms of the one-point function $\langle\varphi_a(x)\rangle=\sqrt{N}\phi_a(x)$ and the symmetric connected -- so-called statistical -- two-point correlator $F_{ab}(x,x')={1\over2}\big<\varphi_a(x)\varphi_b(x')+\varphi_b(x')\varphi_a(x)\big>_c$, with $F(x,x')\equiv F_{aa}(x,x')/N$. The brackets represent an average in an appropriate quantum state.


We wish to investigate dS-invariant solutions of the field equations, for which the one-point function $\phi_a$ is a constant and the two-point function $ F(x,x')=\bar F(s)$ only depends on the dS distance
$ s = (\eta^2+\eta^{\prime2}-|\vec x-\vec x\,'|^2)/2\eta\eta'$. In particular, $F(x,x) = \bar F(1)$, which implies  that the self-consistent mass \eqn{eq:mass} is constant. 

The statistical two-point function can be written in terms of an infinite set of complex mode functions $f_k(\eta)$ in comoving momentum space:
\beq
 F(x,x')=\int\frac{d^dk}{(2\pi)^d}e^{i\vec k\cdot(\vec x-\vec x')}{\rm Re}\left\{f_k(\eta)f^*_k(\eta')\right\}.
\eeq
Using \Eqn{eq:field}, one finds that the rescaled functions $u_k(\eta)=a(\eta)^{d-1\over2}f_k(\eta)$, satisfy the differential equation
\beq
\label{eq:mode}
 u_k''+\left(k^2-\frac{\nu^2-1/4}{\eta^2}\right)u_k=0,
\eeq
where $\nu=\sqrt{{d^2/4}-{M^2}/{H^2}}$.
dS-invariant solutions read \cite{Allen:1985ux}: $u_k(\eta)=\cosh\alpha\, S_k(\eta) + \sinh\alpha \,e^{i\beta} S_k^\star(\eta)$, where
\beq
 S_k(\eta)={1\over2}\sqrt{-\pi\eta}e^{i{\pi\over2}\left(\nu+{1\over2}\right)}H_\nu^{(1)}(-k\eta),
\eeq
with $H_\nu^{(1)}(z)$ the Hankel function, and where the real coefficients $\alpha$ and $\beta$ are $k$ independent.

The requirement of renormalizability further constrains the possible states of the system.
The two-point function in the coincidence limit $F(x,x)$ is divergent and \Eqn{eq:mass} needs to be regularized and renormalized. In principle a dS-invariant regularization (such as e.g. dimensional regularization) should be used. Here, it is sufficient to use a simple ultraviolet (UV) cutoff on physical momenta: $k/a(\eta)=-kH\eta\le\Lambda_p$. Demanding that the short-distance (UV) properties of the theory be independent of the ST geometry, one is left with the Bunch-Davies vacuum, $\alpha=0$, as a unique possibility \cite{Bunch:1978yq,Allen:1985ux,Anderson:2005hi}.

The resummed square mass reads (we make the field dependence explicit from now on)
\beq
\label{eq:mass2}
 M^2(\phi^2)=m^2+d(d+1)\xi H^2+\frac{\lambda}{6}\phi^2+\frac{\lambda}{6}F(x,x),
\eeq
with, writing $\bar\nu={\rm Im}(\nu)$,
\beq
\label{eq:loop}
 F(x,x)=\frac{\sqrt{\pi}}{4\Gamma({d/2})}\!\left(\frac{H}{2\!\sqrt{\pi}}\right)^{\!d-1}\!\!\int_0^\Lambda \!\!dz\, z^{d-1}e^{-\pi\bar\nu}|H_\nu^{(1)}(z)|^2.
\eeq
We see that this is indeed constant if one uses a physical UV cutoff $\Lambda_p=H\Lambda$ \cite{Boyanovsky:1997cr}. A comoving cutoff does not lead to a dS-invariant solution. We shall see below that this is also crucial to recovering the correct Minkowski limit as well as  for a proper renormalization procedure, involving time-independent bare parameters in \Eqn{eq:action}. 

Now the key observation of the present Letter. It follows from the $z\to0$ behavior:
\beq
\label{eqn:IR}
 |H^{(1)}_\nu(z)|^2\sim z^{-2{\rm Re}(\nu)},
\eeq
that the loop integral \eqn{eq:loop} is only defined if ${\rm Re}(\nu)<d/2$, i.e. the only possible solutions to \Eqn{eq:mass2} are such that $M^2(\phi^2)>0$. We emphasize that this is true in any dimension and only follows from the requirement of dS invariance -- it is easily checked to hold for any $\alpha$-vacuum. Thus we see that  the case of a massless minimally coupled field, $M^2=0$, is not allowed: Whatever the value of the tree-level mass, a strictly positive curvature-induced mass squared is generated by the field self-interactions. We conclude that, in the large-$N$ limit, there exits no dS-invariant state for an interacting massless minimally coupled field \footnote{It is well-known that there exists no dS-invariant state for a non-interacting massless minimally coupled field \cite{Allen:1985ux}. The present result concerns the interacting case.}. 

Let us now examine the consequences of this result. Observe first that the field equation
\beq
M^2(\phi^2)\phi_a=0
\eeq
has for unique solution $\phi_a=0$. The strongly enhanced dS quantum fluctuations in the far IR, see \eqn{eqn:IR}, prevent the possible existence of a spontaneously broken phase in any dimension. This is similar to what happens in flat ST in low dimensions, $d<2$ \cite{Coleman:1974jh}. As we shall see shortly, this further implies that any initial symmetry breaking potential is strongly renormalized by ultralong-wavelength fluctuations, eventually leading to radiative symmetry restoration. 
To discuss these issues in more detail, we shall compute the effective potential $V_{\rm eff}(\phi)$. Recalling that $\partial V_{\rm eff}/\partial\phi_a=M^2(\phi^2)\phi_a$ and exploiting the $O(N)$ symmetry, we have, up to an irrelevant constant~$V_0$:
\beq
\label{eq:pot}
 V_{\rm eff}(\phi)=V_0+ {1\over2}\int_0^{\phi^2}\!\!dvM^2(v).
\eeq
Using similar arguments as employed in \cite{Reinosa:2011ut}, it can be shown that $V_{\rm eff}(\phi)$ is convex everywhere.


Before we proceed to actually solving \Eqn{eq:mass2}, we check that we recover the correct flat ST limit $H\to0$, to be taken at fixed $\Lambda_p=\Lambda H$. Note that this also gives the relevant equation for the large mass case $M^2/H^2\gg1$, for which $\nu=i\bar\nu\approx iM/H$. Using the $z\ll\bar\nu$ and $z\gg\bar\nu$ behaviors $|H_{i\bar\nu}(z)|^2\sim\frac{2}{\pi\bar\nu}e^{-\pi\bar\nu}$ and $|H_{i\bar\nu}(z)|^2\sim{2e^{\pi\bar\nu}}/{\pi}\sqrt{z^2+\bar\nu^2}$ respectively, one recovers the large-$N$ mass equation in the Minkowski vacuum \cite{Coleman:1974jh}:
\beq
\label{eq:mink}
 M^2(\phi^2)=m^2+\frac{\lambda}{6}\phi^2+\frac{\lambda}{6}\int\!\frac{d^dp}{(2\pi)^d}\frac{\theta(\Lambda_p-p)}{2\sqrt{p^2+M^2(\phi^2)}},
\eeq
where the contribution from super-Hubble modes is suppressed by a Boltzmann-like factor $e^{-M/T}$ with $T=H/2\pi$ the
temperature associated with the dS horizon.


We now discuss renormalization and specify to $d=3$ spatial dimensions, where the theory is perturbatively renormalizable \footnote{Recall that the present theory has a Landau pole in flat ST in $d=3$ and is only well-defined for a finite UV cutoff unless $\lambda_R\to 0$ \cite{Coleman:1974jh, Reinosa:2011ut}. Renormalization is to be understood as a minimization of the cutoff dependence.}. Using the large-$|z|$ asymptotic expansion $ |H^{(1)}_\nu(z)|^2\sim\frac{2}{\pi z}\left[1+(\nu^2-1/4)/2z^2+{\cal O}(z^{-4})\right]$, one gets the UV-behavior of the loop integral \eqn{eq:loop}:
\beq
\label{eq:div}
  \pi\int_0^{\Lambda}\!\!dz\,z^2|H^{(1)}_\nu(z)|^2=\Lambda^2-\frac{M^2}{H^2}\ln\Lambda+2\ln\Lambda+{\rm finite}
\eeq
where ``finite'' refers to terms which converge in the limit $\Lambda\to\infty$. The first two terms on the right-hand side are just the usual Minkowski UV divergences, see e.g. \Eqn{eq:mink}, whereas the third term is specific to the curved ST. 
The former are eliminated by standard coupling and mass renormalization \cite{Coleman:1974jh}, e.g. by introducing the renormalized mass and coupling at the arbitrary physical scale $\mu$: ${1}/{\lambda_R(\mu)}={1}/{\lambda}+{\ln(\Lambda_p/\mu)}/{48\pi^2}$ and ${m_R^2(\mu)}/{\lambda_R(\mu)}={m^2}/{\lambda}+{\Lambda^2_p}/{48\pi^2}$. The remaining logarithmic divergence can be eliminated by means of \cite{Boyanovsky:1997cr} ${12\xi_R(\mu)}/{\lambda_R(\mu)}={12\xi}/{\lambda}+{\ln(\Lambda_p/\mu)}/{24\pi^2}$. In the following we write the renormalized parameters at the scale $\mu=H$ as $m_R(H)=m_R$ and so on. 
One has the renormalization group (RG)  running:
\beq
\label{eq:run}
 {1\over\lambda_R(\mu)}={1\over\lambda_R}+ \frac{\ln(H/\mu)}{48\pi^2}
\eeq
whereas both $m_R^2(\mu)/\lambda_R(\mu)$ and $\left(\xi_R(\mu)-{1\over6}\right)/\lambda_R(\mu)$ are RG invariants. Equivalently, $(\mu_R^2(\mu)-2H^2)/\lambda_R(\mu)$, where $\mu_R^2(\mu)=m_R^2(\mu)+12\xi_R(\mu) H^2$, is RG invariant. 


It is interesting to note that the conformal case $\nu=1/2$, or $M^2=2H^2$, exhibits an additional symmetry which allows one to scale away the cosmic expansion, see \Eqn{eq:mode}, and which protects the mass from radiative corrections as is easily checked using $|H_{1/2}^{(1)}(z)|^2={2}/{\pi z}$. For instance, at $\phi^2=0$, one gets $M^2(0)=\mu_R^2=2H^2$.


In the limit $M^2/H^2\ll1$, \Eqn{eq:mass2} can be solved analytically. For this purpose, we introduce the small parameter $\epsilon=3/2-\nu\approx M^2/3H^2\ll1$ and separate the momentum integration in \Eqn{eq:loop} in three parts: $\int_0^\Lambda=\int_0^{\kappa'}+\int_{\kappa'}^\kappa+\int_\kappa^\Lambda$ with $\kappa'\ll1\ll\kappa\ll\Lambda_p$. The high-momentum part is evaluated by using the large-$|z|$ asymptotic expansion of $|H^{(1)}_\nu(z)|^2$. We get, up to the UV divergent terms exhibited in \Eqn{eq:div},
\beq
 \int_\kappa^\Lambda \!\!dz\,z^2|H^{(1)}_\nu(z)|^2={\rm div.}-{2\over\pi}\left[\ln{\kappa}+{\kappa^2\over2}+{\cal O}\left(\epsilon,\kappa^{-2}\right)\right].
\eeq
To evaluate the intermediate-momentum contribution we simply set \cite{Boyanovsky:2005px} $\nu\to3/2$ and use $|H^{(1)}_{3/2}(z)|^2={2\over\pi z}(1+z^2)$:
\beq
 \int_{\kappa'}^\kappa\!\!dz\,z^2|H^{(1)}_\nu(z)|^2=\frac{2}{\pi}\left[\ln\frac{\kappa}{\kappa'}+\frac{\kappa^2}{2}+{\cal O}(\epsilon,\kappa^{\prime 2})\right].
\eeq
Finally, the IR contribution is obtained from the $z\to0$ expansion $ |H^{(1)}_\nu(z)|^2\sim\left({2^\nu\Gamma(\nu)}/{\pi z^\nu}\right)^2\left[1+{\cal O}\left(z^{2}\right)\right]$:
\beq
 \int_0^{\kappa'}\!\!dz\,z^2|H^{(1)}_\nu(z)|^2=\frac{2}{\pi}\left[\frac{1}{2\epsilon}-c_3+\ln\kappa'+{\cal O}(\epsilon,\kappa^{\prime2})\right],
\eeq
where $c_3=2-\gamma-\ln2$, with $\gamma$ the Euler constant. In terms of renormalized parameters at the scale $\mu=H$ introduced before, we thus get the following equation:
\beq
\label{eq:masseq}
 M^2(\phi^2)=\mu_R^2+\frac{\lambda_R}{6}\phi^2+\frac{\lambda_RH^2}{48\pi^2}\left[\frac{3H^2}{M^2(\phi^2)}-2\bar c_3\right],
\eeq
where $\bar c_3 = c_3 + 1/6 \approx 0.9$. The positive solution is \footnote{A similar result was obtained at $\phi^2=0$ for the $N=1$ case in the Hartree approximation in \Ref{Garbrecht:2011gu}.}
\beq
\label{eq:massapp}
 M^2(\phi^2)=a(\phi^2)+\sqrt{a^2(\phi^2)+b}
\eeq
with
\beq
 a(\phi^2)=\frac{\mu_R^2}{2}-\frac{\bar c_3\lambda_R}{48\pi^2}H^2+\frac{\lambda_R}{12}\phi^2, \,{\rm and}\,\,\, b=\frac{\lambda_RH^4}{16\pi^2}
\eeq
The nonanalytic $\lambda_R$-dependence is typical of IR effects.

In this approximation, the effective potential \eqn{eq:pot} can be analytically integrated and reads (we choose $V_0=0$):
\beq
\label{eq:potapp}
 V_{\rm eff}(\phi)=\frac{3}{2\lambda_R}\left(M^4(\phi^2)-M^2(0)\right)+\frac{3H^4}{16\pi^2}\ln\frac{M^2(\phi^2)}{M^2(0)}.
\eeq
Full numerical solutions of Eqs. \eqn{eq:mass2} and \eqn{eq:pot} reveal that, for a wide range of couplings, Eqs.~\eqn{eq:massapp}-\eqn{eq:potapp} are in fact quantitatively accurate even for nonsmall values of $M^2/H^2$. This is because when the infrared term $H^2/M^2$ on the right-hand side of \Eqn{eq:masseq} is not numerically important, the loop corrections are typically suppressed by a factor $\lambda_R/48\pi^2$. Below, we use \Eqn{eq:massapp} to illustrate our findings, all of which we checked numerically.


The above result shows the phenomenon of curvature-induced mass generation \cite{Starobinsky:1994bd}. Indeed, consider the case $\mu_R^2+\lambda_R\phi^2/6=0$ \footnote{Note that this condition is not RG invariant, see \Eqn{eq:run}.}, which includes the tree-level massless minimally coupled field at $\phi^2=0$, relevant for inflation physics, or the Goldstone mode of a tree-level potential exhibiting spontaneous symmetry breaking, i.e. with $\mu_R^2<0$. In this case, the perturbative expansion is ill defined due to IR divergences. Self-interactions screen these IR effects and generate a nonvanishing effective mass and/or nonminimal coupling:
\beq
 \frac{M^2}{H^2}=-\frac{\bar c_3\lambda_R}{48\pi^2}+\sqrt{\frac{\lambda_R}{16\pi^2}+\left(\frac{\bar c_3\lambda_R}{48\pi^2}\right)^2}.
\eeq
For weak coupling, where the approximation \eqn{eq:masseq} makes sense, we get $M^2/H^2=\sqrt{\lambda_R}/4\pi-\bar c_3\lambda_R/48\pi^2+{\cal O}(\lambda_R^{3/2})$. The first term is the Starobinsky-Yokoyama result \cite{Starobinsky:1994bd}, obtained here directly from the basic QFT, see also \cite{Burgess:2009bs}. As discussed in \cite{Sloth:2006az, Riotto:2008mv} this curvature generated mass leads to an enhancement factor $H^2/M^2\propto1/\sqrt{\lambda_R}$ of IR-sensitive observables, such as parametrically ${\cal O}(\sqrt{\lambda_R})$ non-Gaussianities or corrections to the power spectrum.


\begin{figure}[t!]
 \centerline{\epsfxsize=6.5cm\epsffile{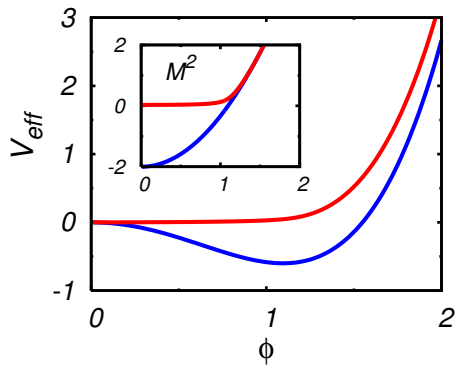}\vspace*{-0.2cm}}
 \caption{\label{fig:Veff} The effective (red) vs tree-level (blue) potentials as a function of $\phi$ for $\mu_R^2=-2$ and $\lambda_R=10$, all in units of $H$. The inset shows the square mass $M^2(\phi^2)$ (red) and its tree-level expression $\mu_R^2+\lambda_R\phi^2/6$ (blue). Here, the numerical result agrees with Eqs. \eqn{eq:massapp}-\eqn{eq:potapp} over the whole range of $\phi$.
 \vspace*{-0.4cm}}
\end{figure}

Equation \eqn{eq:massapp} also illustrates the phenomenon of radiative symmetry restoration for $\mu_R^2<0$. In that case, one gets, at $\phi^2=0$, $M^2/H^2=\lambda_RH^2/16\pi^2|\mu_R^2|+{\cal O}(\lambda_R^2)$. Notice the parametric $\lambda_R$ dependence which may lead to an even stronger enhancement of IR-sensitive observables than mentioned above. As an illustration, \Fig{fig:Veff} shows an exact numerical solution of Eqs. \eqn{eq:mass2} and \eqn{eq:pot}. 


In conclusion, we find that the strongly enhanced ultralong-wavelength scalar field fluctuations in dS prevent IR divergences by generating a strictly positive curvature-induced square mass and, as a consequence, forbid the possible existence of a spontaneously broken phase. This is a genuine nonperturbative effect. It is an interesting question to investigate whether this result remains true beyond the large-N limit \cite{PS}. Together with Refs. \cite{Riotto:2008mv, Burgess:2009bs} this work provides a step toward the precise understanding of such nonperturbative effects in dS from first principles.

We emphasize that the phenomenological relevance of these results requires a rather long period of (nearly) dS expansion in order for such IR fluctuations to develop. In the small mass limit \eqn{eq:masseq}, the relevant wavelength are as large as $\exp(3H^2/2M^2)$. This is certainly not a problem for e.g. eternal inflation scenarios.

The phenomena of mass generation or symmetry restoration are similar to finite temperature effects in flat ST. It is a very interesting question whether other important thermal effects such as damping of unequal time correlators \cite{Berges:2004vw} occur in dS. This requires calculations beyond the present approximation \cite{Garbrecht:2011gu,PS}.

Various techniques have been developed to deal with IR and secular issues in equilibrium and nonequilibrium QFT in flat ST \cite{Blaizot:2001nr, Boyanovsky:1998aa, Berges:2004vw}. It remains to be investigated whether these can help resolving similar questions in curved geometries, or if specific methods will be needed.


We acknowledge useful discussions with R. Parentani and J. Pawlowski.






\end{document}